\documentclass[aps,prb,preprint,showpacs,unsortedaddress,endfloats]{revtex4}
\usepackage{graphicx}
\usepackage{setspace}
\usepackage{bm}
\usepackage{amsfonts}
\usepackage{amsmath}
\usepackage{amssymb}%
\newcommand{\be}{\begin{equation}}
\newcommand{\ee}{\end{equation}}
\newcommand{\bd}{\begin{displaymath}}
\newcommand{\ed}{\end{displaymath}}
\newcommand{\ba}{\begin{array}}
\newcommand{\ea}{\end{array}}
\newcommand{\bq}{\begin{eqnarray}}
\newcommand{\eq}{\end{eqnarray}}

\begin{document}
\begin{spacing}{2}
\title{Diffraction in low-energy electron scattering from DNA: bridging gas phase and solid state theory}

\author{Laurent Caron}
\affiliation{Groupe de Recherches en Sciences des Radiations,
Facult\'{e} de m\'{e}decine, Universit\'{e} de Sherbrooke,
Sherbrooke, QC J1H 5N4, Canada.}
\altaffiliation[Permanent address: ]{D\'{e}partement de physique et Regroupement qu\'{e}b\'{e}cois sur les
mat\'{e}riaux de pointe, Universit\'{e} de Sherbrooke, Sherbrooke, QC J1K 2R1, Canada.}
\author{Stefano Tonzani}
\affiliation{Department of Chemistry,  Northwestern University, Evanston, IL 60208-3113}
\email[Corresponding author: ]{tonzani@northwestern.edu}
\author{Chris H. Greene}
\affiliation{Department of Physics and JILA, University of Colorado, Boulder, Colorado 80309-0440}
\author{L\'{e}on Sanche}
\affiliation{Groupe de Recherches en Sciences des Radiations,
Facult\'{e} de m\'{e}decine, Universit\'{e} de Sherbrooke,
Sherbrooke, QC J1H 5N4, Canada.}
\date{\today}

\begin{abstract}
Using high-quality gas phase electron scattering calculations and multiple
scattering theory, we attempt to gain insights on the radiation damage to DNA
induced by secondary low-energy electrons in the condensed phase, and to bridge
the existing gap with the gas phase theory and experiments. The origin of 
different resonant features (arising from single molecules or diffraction) 
is discussed and the calculations are compared to existing
experiments in thin films.
\end{abstract}

\maketitle

\section{Introduction}
Since it was discovered \cite{Sanche:DNA} that low
energy electrons (LEE) can cause strand breaks in the DNA duplex, 
the interest in electron interactions with DNA, which originates prevalently from
its importance in radiation damage to living tissue and radiotherapy, has grown
consistently \cite{Sanche:EurPhys2005}. The impact of ionizing electromagnetic radiation on matter causes 
the emission of highly energetic electrons; these latter ionize the medium via 
an electromagnetic type of interaction, thus producing secondary electrons in 
large numbers\cite{Cobut98}. Most secondary electrons are created with low energy 
(E$<$20 eV) and have a distribution with a most probable energy of 9 eV \cite{Pimblott:2007}. 
If the electron energy is above the DNA ionization threshold (7-10 eV) these 
electrons can ionize it, while over the entire 0-15 eV range they can be captured 
in a resonant anionic state. Fast, efficient dissociation pathways are known to 
exist for organic molecules \cite{Bass98} from electronic dissociative states, 
dissociative electron attachment, dipolar dissociation and dissociative ionization, 
which can be related to capture or transfer of an electron or hole into a dissociative
state.\cite{Barrios:JPCB02,Burrow:PRL04} 

In the context of sub-ionization threshold electrons, the importance of
resonances has become evident. Specifically, electron capture by a DNA 
subunit to form a molecular resonance enhances greatly the rupture of 
chemical bonds within the molecule either by dissociative electron 
attachment (DEA) or the decay of the transient anion into a dissociative 
electronically excited state.  These phenomena are reflected in the 
measured yield of strand breaks which exhibit prominent resonant 
features as a function of energy \cite{sanche2003}.

In a condensed environment, LEE are created within or outside DNA.  The 
former necessarily have a high probability to interact with DNA.  LEE 
created outside DNA may also interact with this molecule depending on 
the timescale of thermalization of the electrons.  In this case, 
other types of damage, such as radical-mediated damage, can
become important or even predominant. In general, if the thermalization is 
slow, DNA is impacted by a relatively "hot" electron (meaning in
this context an electron not fully solvated) from outside; this entails usually a timescale
between 0.1 and 100 fs.\cite{Migus:PRL87} Others created outside the DNA do not
arrive at the target with any appreciable energy and therefore cannot directly
damage it, but they can form radicals along the way and these in turn can
attack the DNA \cite{VonSonntag:AQC07}. The prevalence of one or the other of
these mechanisms probably is determined by the DNA concentration: if it is
high, then the electrons are more likely to be created within DNA or impact it before being completely
solvated. In principle, modeling the dynamics of the radiation tracks\cite{Uehara} can yield
very useful informations in this regard.

Since radiation damage in a cell is a complicated problem, researchers have
tried to simplify its description by observing first what happens to DNA
components in the gas phase when impacted by LEEs; this literature is by now rather vast. 
The experimental approaches have been focused mostly on predicting
products of dissociative electron attachment from DNA subunits and determining
the role of shape resonances
\cite{Denifl2004,Scheier:AngChem06,Burrow98,Sanche_Burrow:PRL04}. 
The theoretical community has provided information on resonances
in elastic and inelastic scattering
\cite{Tonzani:JCP05,Gorfinkiel:JPB06,Gianturco:PRL04,Trevisan_Orel_THF:06,McKoy_C60:PRA06}.
While dissociative electron attachment
calculations for these large targets are still beyond reach, quantum
chemical methods have been employed to predict the weakest bonds in electron
attachment \cite{Barrios:JPCB02,Schaefer:JACS05} and study the hydration effects on the anionic
compounds \cite{Schaefer:JCP06}. It must be noted, however, how none of these
methods will be able for the foreseeable future to deal with a molecule of tens of
thousands of atoms like bacterial DNA in a process in which both electronic and nuclear
degrees of freedom are involved, the scattering electron is unbound and the
solvent has to be considered as well. While DFT is able to deal with large
systems, nuclear dynamics on DFT surfaces is not as well defined
\cite{Prezhdo:PRL05,Maitra:JCP06} and a scattering theory using DFT is just
starting to be developed
\cite{Burke_Wasserman:JCP05}.

This work aims to be a first step towards systematizing what happens in passing
from the gas to the condensed phase. In particular, we will be concerned with
the solid state, since  this gives us the opportunity to relate to radiation
damage experiments in thin
film \cite{Sanche:DNA,Sanche:EurPhys2005} and, equally importantly,  to neglect fluctuations in
the DNA structure, which are of
fundamental importance in solution \cite{Lewis2003,Lewis_Schatz_Long:JACS05}.

To this extent, the first step is to understand what happens inside the DNA
polymer itself, which is represented here as rigid and immersed in vacuum. In
this context it is possible to explore parameters like coherence
length (which controls the exponential decay of the amplitude of the phase
coherent part of the electron wave function), DNA sequence, and conformation, which
could be important in radiobiological damage.

We pursue this goal using a model, described below, which unites a
recently developed multiple scattering framework \cite{PRL,PRA,PRA2,PRA3} with
accurate electron scattering calculations for the DNA bases performed using the R-matrix
method \cite{Tonzani:JCP05,Tonzani:JCP06}.

In our model we have left out of the picture, for the moment, all the properties
of the liquid phase, from fluctuations to the motion of the solvated electron
in water.
We plan to introduce the structural water (also present in thin films) as a scattering dopant in further
studies using this model, which will be fairly straightforward. To introduce the liquid
phase properties\cite{Chandler:ARPC94} would be instead much more difficult and require some essential
modification to this framework. 
 
\section{Model}
The multiple scattering framework we shall be using to get information on the
elastic scattering of low-energy electrons (LEE) has been described in the
series of articles by Caron and Sanche \cite{PRL,PRA,PRA2,PRA3}. It was
developed at a time when no scattering matrix information was available for
the DNA bases. The objective of the toy model used was to find trends in cross
sections and capture amplitudes resulting from regularity or disorder in the
helix and base sequencing. With the advent of recently available scattering
information on DNA bases, it is important to revisit elastic LEE scattering on
DNA to get more precise numerical estimates and ascertain whether or not the
previous conclusions are still valid. One obvious difference of the current
calculations with respect to Refs. \onlinecite{PRL,PRA,PRA2,PRA3} is the presence of
shape resonances that carry over from the gas phase scattering of the subunits,
which makes it possible to observe the fate of local resonances in the
conjugate.
We shall first study, in this paper, the
idealized B-form of a GCGAATTGGC decamer (without backbone) \cite{ProtExp} and its
regularly sequenced cousin, the $ \text{poly(A)} \cdot \text{poly(T)}$ base
pairs decamer. We shall then examine the A-form of the GCGAATTGGC decamer \cite{ProtExp}. These structures are chosen to be ideal ones, although they are
in general known to be slightly different\cite{Pettitt:BJ98}. Our prototype
systems are decamers because 10 is the number of base pairs in one turn of the helix of B-form DNA.

But let us first review the theoretical framework.

\subsection{Multiple Scattering Theory}
\label{sec:mult_scat}
In Refs \onlinecite{PRL} and \onlinecite{PRA}, we presented the basic equations for
multiple electron scattering within macromolecules, including DNA. For the
latter, we proposed a simple model of molecular subunits (i.e. bases, sugars,
and phosphates) immersed in an optical potential $U_{op}$, which is constant
between their R-matrix shells (or between the muffin-tins), a working
hypothesis that has been used in the calculations for simple molecules
\cite{DillDehmer}, in the theory of low-energy electron diffraction (LEED) in
solids \cite{pendry} and nanoscale structures\cite{Heller:RMP03} . The only function
of the real part of the optical potential is to account for the average energy seen
by an electron. One can quite generally describe the scattering problem
of a molecular subunit by its scattering matrix $S_{L^{\prime}L}$
\cite{MottMassey,gianturco1986} where $L=(l,m)$ are the angular momentum
quantum numbers. Each molecular subunit has an incident plane wave of momentum
$\vec{k}$ impinging on it plus the scattered waves of all other subunits. More
specifically, we described the asymptotic form of the total wave function
$\psi_{\vec{k}}^{(n)}(\vec{r})$ for a molecule centered at $\vec{R}_{n}%
$\ outside the R-matrix shell by the following equation%
\begin{equation}
\psi_{\vec{k}}^{(n)}(\vec{r})=4\pi e^{i\vec{k}\cdot\vec{R}_{n}}\sum
_{LL^{\prime}}i^{l}B_{\vec{k}L}^{(n)}Y_{L^{\prime}}\left(  \Omega_{\vec{r}%
_{n}}\right)  \left[  j_{l}\left(  kr_{n}\right)  \delta_{L^{\prime}L}%
+\frac{1}{2}\left(  S_{L^{\prime}L}^{(n)}-\delta_{L^{\prime}L}\right)
h_{l^{\prime}}^{(1)}\left(  kr_{n}\right)  \right]  \,, \label{eq1}%
\end{equation}
where $Y_{L}$ are spherical harmonics, $j_{l}$ and $h_{l^{\prime}}^{(1)}$ are
the spherical Bessel function and Hankel function of the first kind
respectively, $\vec{r}_{n}=\vec{r}-\vec{R}_{n}$, and%
\begin{align}
B_{\vec{k}L}^{(n)}  &  =Y_{L}^{\ast}\left(  \Omega_{\vec{k}}\right)  +\frac
{1}{2}\sum_{n^{\prime}\neq n}\sum_{L_{1},L_{2},L_{2}^{\prime}}i^{l_{1}%
+l_{2}-l_{2}^{\prime}}B_{\vec{k}L_{2}}^{(n^{\prime})}\left(  S_{L_{2}^{\prime
}L_{2}}^{(n^{\prime})}-\delta_{L_{2}^{\prime}L_{2}}\right) \label{eqelastic}\\
&  \times(-1)^{m_{2}^{\prime}}e^{-i\vec{k}\cdot\vec{R}_{nn^{\prime}}}%
F_{m_{1},m,-m_{2}^{\prime}}^{l_{1},l,l_{2}^{\prime}}Y_{L_{1}}\left(
\Omega_{\vec{R}_{nn^{\prime}}}\right)  h_{l_{1}}^{(1)}\left(  kR_{nn^{\prime}%
}\right) \nonumber
\end{align}
where
\begin{align*}
F_{m_{1},m,-m_{2}^{\prime}}^{l_{1},l,l_{2}^{\prime}}  &  =\left[  4\pi(2l_{1}%
+1)(2l+1)(2l^{\prime}_{2}+1)\right]  ^{\frac{1}{2}}\\
&  \times\left(
\genfrac{}{}{0pt}{}{l_{1}}{0}%
\genfrac{}{}{0pt}{}{l}{0}%
\genfrac{}{}{0pt}{}{l_{2}^{\prime}}{0}%
\right)  \left(
\genfrac{}{}{0pt}{}{l_{1}}{m_{1}}%
\genfrac{}{}{0pt}{}{l}{m}%
\genfrac{}{}{0pt}{}{l_{2}}{-m^{\prime}_{2}}%
\right)  \text{ ,}%
\end{align*}
and $\left(
\genfrac{}{}{0pt}{}{l_{1}}{m_{1}}%
\genfrac{}{}{0pt}{}{l}{m}%
\genfrac{}{}{0pt}{}{l^{\prime}_{2}}{-m^{\prime}_{2}}%
\right)  $ is the Wigner 3-j symbol \cite{messiah}, and $\vec{R}_{nn^{\prime}%
}=\vec{R}_{n}-\vec{R}_{n^{\prime}}$. Equation (\ref{eqelastic}) implies a
coupled set of linear equations for all $B_{\overrightarrow{k}L}^{(n)}$, which
measure the resultant of the superposition of the incident plane wave and the
contribution from all other scatterers. As mentioned before \cite{PRL,PRA},
the loss of coherence of the electrons due to inelastic collisions and perhaps also to the
presence of parasite scatterers (e.g. the water molecules in the grooves could
be considered as such) can be invoked through an imaginary part in the
background optical potential $U_{op}$ \cite{pendry}, i.e. an imaginary part to
the electron wave number $\text{Im}(k)=\xi^{-1}$. Here $\xi$ acts as a
coherence length for the electrons.

\subsection{Electron capture and scattering}

In an effort to extract physically meaningful information from the multiple
scattering formalism, we had previously targeted a calculation of the capture
amplitude $V_{\vec{k}}^{(n)}$ of an electron in a shape or core excited
resonance of a basic subunit positioned at $\vec{R}_{n}$. We had assumed a
dominant capture channel symmetry corresponding to $L_{o}$ and had used the
one-center approximation of O'Malley and Taylor \cite{OMalley1968}. When generalized to a
multiple scattering situation, this leads to%
\begin{equation}
V_{\vec{k}}^{(n)}=\sqrt{4\pi}V_{L_{o}}B_{\vec{k}L_{o}}^{(n)}e^{i\vec{k}%
\cdot\vec{R}_{n}}\;, \label{eqV1}%
\end{equation}
where $V_{L_{o}}$ is an energy and nuclear coordinate dependent amplitude.
There is unfortunately no available theoretical information on the nuclear
part of the wavefunction for the DNA bases at this time. So we shall only focus on the electronic part.

We proposed \cite{PRA2}\ a weighted partial capture factor%
\begin{equation}
\Gamma_{w}(l_{o})=\frac{\sum_{\vec{R}_{n}}\gamma(l_{o},\vec{R}_{n})}%
{\sum_{\vec{R}_{n}}}~, \label{gamma2}%
\end{equation}
where the constituent partial capture factor%
\begin{equation}
\gamma(l_{o},\vec{R}_{n})=\frac{\sum_{m_{o}=-l}^{l}\left\vert \sqrt{4\pi
}B_{\vec{k}l_{o}m_{o}}^{(n)}\right\vert ^{2}}{(2l+1)}~, \label{gamma1}%
\end{equation}
measures the partial wave decomposition of the total wave function averaged
over the different bases positioned at $\vec{R}_{n}$. This would serve as meaningful measure
of the effect of multiple scattering on the capture probability in the $l_{o}$
channel since $\gamma(l_{o},\vec{R}_{n})=1$ for a lone plane wave. Note that
$\gamma(0,\vec{R}_{n})$ equals the average absolute square of the wave
function at $\vec{R}_{n}$ and thus $\Gamma_{w}(0)$ measures the absolute square of the wave
function averaged over all bases.

The total elastic cross section, for a finite size macromolecule, would also
be of interest. This is, however, a somewhat elusive quantity when there are
losses. Technically, we can expand the scattered part of Eq. (\ref{eq1})
around the geometric center $\overrightarrow{R}_{_{GC}}$ of the macromolecule.
In this reference system, remembering that $\vec{r}_{n}=\vec{r}-\vec{R}_{n}$, one has%
\begin{align}
Y_{L^{\prime}}\left(  \Omega_{\vec{r}_{n}}\right)  h_{l^{\prime}}^{(1)}\left(
kr_{n}\right)   &  =\sum_{L_{1},L_{2}}i^{l_{1}+l_{2}-l^{\prime}}%
(-1)^{m^{\prime}}F_{m_{1,}m_{2,}-m^{\prime}}^{l_{1},l_{2},l^{\prime}}Y_{L_{1}%
}\left(  \Omega_{\overrightarrow{r}-\overrightarrow{R}_{GC}
}\right)  \\
&  \times Y_{L_{2}}\left(  \Omega_{\overrightarrow{R}_{GC}%
-\overrightarrow{R}_{n}  }\right)  h_{l_{1}}^{(1)}\left(  k\left\vert
\vec{r}-\vec{R}_{GC}\right\vert \right)  j_{l_{2}}\left(
k\left\vert \vec{R}_{GC}-\vec{R}_{n}\right\vert \right)
~.\nonumber
\end{align}
In the limit $\left\vert \vec{r}-\vec{R}_{GC}\right\vert
$ large, one can write
\begin{equation}
\lim_{\rho\rightarrow\infty}h_{l_{1}}^{(1)}\left(  k\rho\right)  =i^{-l_{1}%
-1}e^{ik\rho}/(k\rho) , \label{eq_h1}%
\end{equation}
where $\vec{\rho}=\vec{r}-\vec{R}_{GC}$. Therefore, one
obtains%
\begin{align}
\lim_{\rho\rightarrow\infty}\psi_{\vec{k}}^{(n)}(\vec
{\rho}) &  =2\pi \sum_{LL^{\prime}}e^{i\vec{k}\cdot\vec{R}_{n}}i^{l}
B_{\vec{k}L}^{(n)}T_{L^{\prime}%
L}^{(n)}\sum_{L_{1},L_{2}}i^{l_{1}+l_{2}-l^{\prime}}(-1)^{m^{\prime}}%
F_{m_{1,}m_{2,}-m^{\prime}}^{l_{1},l_{2},l^{\prime}}Y_{L_{1}}\left(
\Omega_{\vec{\rho}}\right)   \label{eq_psin}  \\
&  \times Y_{L_{2}}\left(  \Omega_{\vec
{R}_{GC}-\vec{R}_{n}}\right) j_{l_{2}}\left(%
k\left\vert \vec{R}_{GC}-\vec%
{R}_{n}\right\vert \right)  i^{-l_{1}-1}e^{ik\rho}/(k\rho)~.\nonumber
\end{align}
where%
\begin{equation}
T_{L^{\prime}L}^{(n)}=\left[  S_{L^{\prime}L}^{(n)}-\delta_{L^{\prime}%
L}\right]  ~.
\end{equation}
is the T-matrix. From this, one can calculate the scattered current at
distance $\rho$ divided by the incident electron flux. In doing this, the only part of
the exponential term $e^{ik\rho}$ of Eqs \ref{eq_h1} and \ref{eq_psin} to survive will
be due to the imaginary part of the wave number $\xi^{-1}$. One can write
\begin{equation}
\sigma_{e}(k)=\sigma_{pe}(k)e^{-2\rho/\xi} \label{eq_pseudosigma}
\end{equation}
with the following definition for the pseudo-elastic cross section%
\begin{equation}
\sigma_{pe}(k)=\sum_{L_{1}}\left\vert \phi_{L_{1}}\right\vert ^{2}/\left\vert
k\right\vert ^{2}~,%
\end{equation}%
\begin{align}
\phi_{L_{1}} &  =2\pi\sum_{nLL^{\prime}}e^{i\vec{k}\cdot\vec{R}_{n}}B_{\vec{k}L}^{(n)}%
T_{L^{\prime}L}^{(n)}\sum_{L_{2}}i^{l+l_{2}-l^{\prime}-1}(-1)^{m^{\prime}%
}F_{m_{1,}m_{2,}-m^{\prime}}^{l_{1},l_{2},l^{\prime}}\\
&  \times Y_{L_{2}}\left(  \Omega_{\vec{R}_{GC}%
-\vec{R}_{n}}\right)  j_{l_{2}}\left(  k\left\vert
\vec{R}_{GC}-\vec{R}_{n}\right\vert \right)  .\nonumber
\end{align}
The exponential decay in Eq. (\ref{eq_pseudosigma}) is the result of the loss
of coherence of the electron as it travels and is due to scatterers external to the
decamer. Whereas $\lim_{\rho\rightarrow\infty}\sigma_{e}$ goes to zero for all finite
$\xi$, $\sigma_{pe}$ remains finite. For all practical purposes, this last
quantity tends to the normal definition of cross section $\sigma=\lim
_{\xi\rightarrow\infty}\sigma_{e}$ whenever $\xi$ is much larger than the size
of the macromolecule.

\subsection{R-matrix calculations}

The static-exchange approximation\cite{Morr_Coll:PRA78}
reduces the problem of electron scattering from a polyatomic molecule to  
a one electron problem. This approximation amounts to including only the
ground state of the target in the close coupling expansion of the wave
function, and it is roughly the equivalent of the Hartree-Fock
approximation for continuum states. \cite{Froese-Fischer:book} 
A detailed description of our method can be found in Refs.
\onlinecite{Tonzani:JCP05,Tonzani:JCP06,Tonzani:CPC06, Tonzani:THF}; here we just
sketch the main points of the treatment.

We use
the R-matrix method  to solve the one electron problem. \cite{Tonzani:JCP05}
This method consists in partitioning space into  
a short range zone, where all the channels are coupled and the scattering
problem can in principle  be treated in all its many body complexity, and an outer
zone (external to the target electron density) in which the escaping electron only sees the effect of the target molecule as a
multipole expansion of the electrostatic potential. In passing, we note that
this is conceptually very suitable to the scheme we are applying in this work,
patching together short-range scattering data for various DNA subunits, and
it also allows us to get rid of the long-range (dipole) part of the electron
molecule interactions, which is not relevant here.
In its eigenchannel form, the R-matrix method can be formulated as a
variational principle \cite{Greene:rev96} for the normal logarithmic derivative
(-$b$) of the
wavefunction on the reaction zone surface:
\be
\label{log:der}
{b\equiv - \frac{\partial{\log {(r\Psi)}}}{\partial r} = 2 \frac
{\int_V {\Psi^{*}(E-\hat{H} -\hat{L}) \Psi dV}}{\int_V{\Psi^{*}
\delta(r-r_0)\Psi dV}}}
\ee
where $\hat{L}$ is the Bloch operator, needed to make the Hamiltonian $\hat{H}$
Hermitian  and $r_{o}$ is the
boundary between the internal and external regions.
It is possible, after expanding the internal region wavefunction in a
suitable basis set, to recast the solution of Eq. \ref{log:der}
as an eigenvalue problem and then
through basis set partitioning to shift the computational burden to the
solution of a large linear system. \cite{Tonzani:JCP05, Greene:rev96}
As a basis set we use finite elements \cite{FEMBEM:notes} in all three
spherical coordinates, in this way we have large but sparse matrices that are amenable
for solution with fast sparse solvers. \cite{Tonzani:CPC06,pardiso}

A further simplification consists in using the local density approximation (LDA) for
the exchange potential:
\be
\label{Pot:exch}
V_{ex}(\vec{r}) = -\frac{2}{\pi} k_{F} F(k_{F},E),
\ee
where $k_{F}$ is the local Fermi momentum:
\be
k_{F}(\vec{r})=[3 \pi ^{2} \rho(\vec{r})]^{1/3}
\ee
and $F$ is a functional of the energy and the local
density $\rho(\vec{r})$ (through the local Fermi momentum). The  functional
form we use for $F$ is called the Hara
exchange. \cite{Hara:69}  It has been extensively employed in continuum state
calculations, and it is energy-dependent.  The LDA, widely used also in
density functional (DFT) calculations, \cite{Parr_Yang:book} gives qualitatively correct
results, \cite{Tonzani:JCP05,Morr_Coll:PRA78} while it is simple enough
to allow calculations for complex molecular targets.

A polarization-correlation potential is added to this.
The long range part of this potential is
a simple multipole expansion, of which we retain only the induced dipole
polarization term: 
\be
\label{polar:potential}
V_{pol} = -\frac{\alpha_{0}}{2 r^{4}}
\ee
where $\alpha_{0}$  is
the totally symmetric 
component of the polarizability tensor (higher order and anisotropic terms are much less
important \cite{Tonzani:THF}), and it can be calculated $ab$ 
$initio$ using electronic structure codes.
This potential is in principle nonlocal inside the target molecule. 
We approximate it as a local potential using a form based on DFT
(specifically on the LYP potential of
Ref. \onlinecite{Lee_Yang_Parr:PRB88}) which has yielded
reliable results in the work of Gianturco and coworkers. \cite{Gianturco:c60}
This form makes use of the electron density, its gradient and Laplacian, which have
to be calculated for each target molecule. The short and long range potentials are
matched unambiguously (continuously but with discontinuous derivatives) at the innermost crossing point, whose radius is dependent on the
angles. The matching is unambiguous in the sense that there are two crossing
points between the inner and outer potential for each angle, and we always
choose the innermost, since the other is far in the region where the electron
density of the molecule is very small. Choosing the outermost crossing has proven to give
unphysical results \cite{Gianturco:PRA93} in many cases.

All the target quantities are calculated at the Hartree-Fock level using a
6-31G** basis set, and the target equilibrium geometries have been optimized at the
same level of theory.
The details of the calculations have been
described in Ref. \onlinecite{Tonzani:JCP05}, including the
convergence criteria. We notice here that the cross sections for the purines
had to be recalculated due to an error in the calculation of the electron
 density, as noted in Refs. \onlinecite{Tonzani:THF,Tonzani:PhD_thesis}, and their resonances
are now shifted to lower energy with respect to Ref.
\onlinecite{Tonzani:JCP05}.  These calculations are very cumbersome, and for the level of
accuracy we are aiming for here this  convergence criterion seems adequate.  

\section{Results and Discussion}
\label{sec:results}
\subsection{B-form $\text{poly(A)}\cdot \text{poly(T)}$ base pairs decamer}

We found it judicious to start our study with the very regular $\text{poly(A)}\cdot
\text{poly(T)}$ decamer
(with one strand containing 10 adenines and the other 10 thymines, not
considering the backbone) in order to avoid sequence disorder
and focus more on the regularity of the spiral structure. This should favor
comparisons with the previous toy model simulations which predicted important
effects of internal diffraction and enhancement of capture amplitudes at low
energy \cite{PRL,PRA}. All calculations in this section were done for a
coherence length of $\xi=1000$ a.u. which is roughly 15 times the decamer
length. This is long enough for good coherence throughout the decamer and the
use of $\sigma_{pe}$ as a measure of the total elastic cross section.

We used the atomic coordinates of an AT pair in the idealized structure data
file obtained from the Protein Explorer internet site \cite{ProtExp} and generated the
decamer by rotating this structure by 36 degrees and translating it by 3.38
\AA\ along the spiral axis \cite{Neidle}. Our $S$ matrices were calculated using a
dipole cutoff at 12 a.u., the size of our R-matrix box. The reasoning behind this is that the superposition
of the long-range part coming from all bases would nearly cancel out and, in any case, be
absorbed into the optical potential. The $S$ matrices of
adenine and thymine had to be rotated so that their principal molecular axes
coincided with those of the corresponding bases in the decamer. Three
orthogonal axes were defined for each base from the positions of the C2, C4,
and C6 atoms of the ring part \cite{Neidle}, $\vec{a}_{Ri}$ for the R-matrix
molecule positioning and $\vec{a}_{dj}$ for the decamer base orientation
($i,j=1,2,3$). The following transformation is inspired by Messiah's
\cite{messiah} treatment of rotations. We define three Euler angles $\alpha$,
$\beta$, and $\gamma$ from the following association of%
\begin{equation}
R(\alpha,\beta,\gamma)_{i,j}=\vec{a}_{Ri}\cdot\vec{a}_{dj}%
\end{equation}
with Eq$.$ (C.45) of Messiah. The transformation matrix is then defined as%
\begin{equation}
W_{lm,lm^{\prime}}=e^{-i\alpha m}\tau_{mm^{\prime}}^{(l)}(\beta)e^{-i\gamma
m^{\prime}},
\end{equation}
where $\tau_{mm^{\prime}}^{(l)}(\beta)$ is the $y$ axis rotation matrix
defined in Eq$.$ (C.72) of Messiah. The transformed S matrix $\tilde
{S}_{lm,l^{\prime}m^{\prime}}$ becomes%
\begin{equation}
\tilde{S}=WSW^{-1}.
\end{equation}

\begin{figure}[tbh]
\includegraphics[scale=.8]{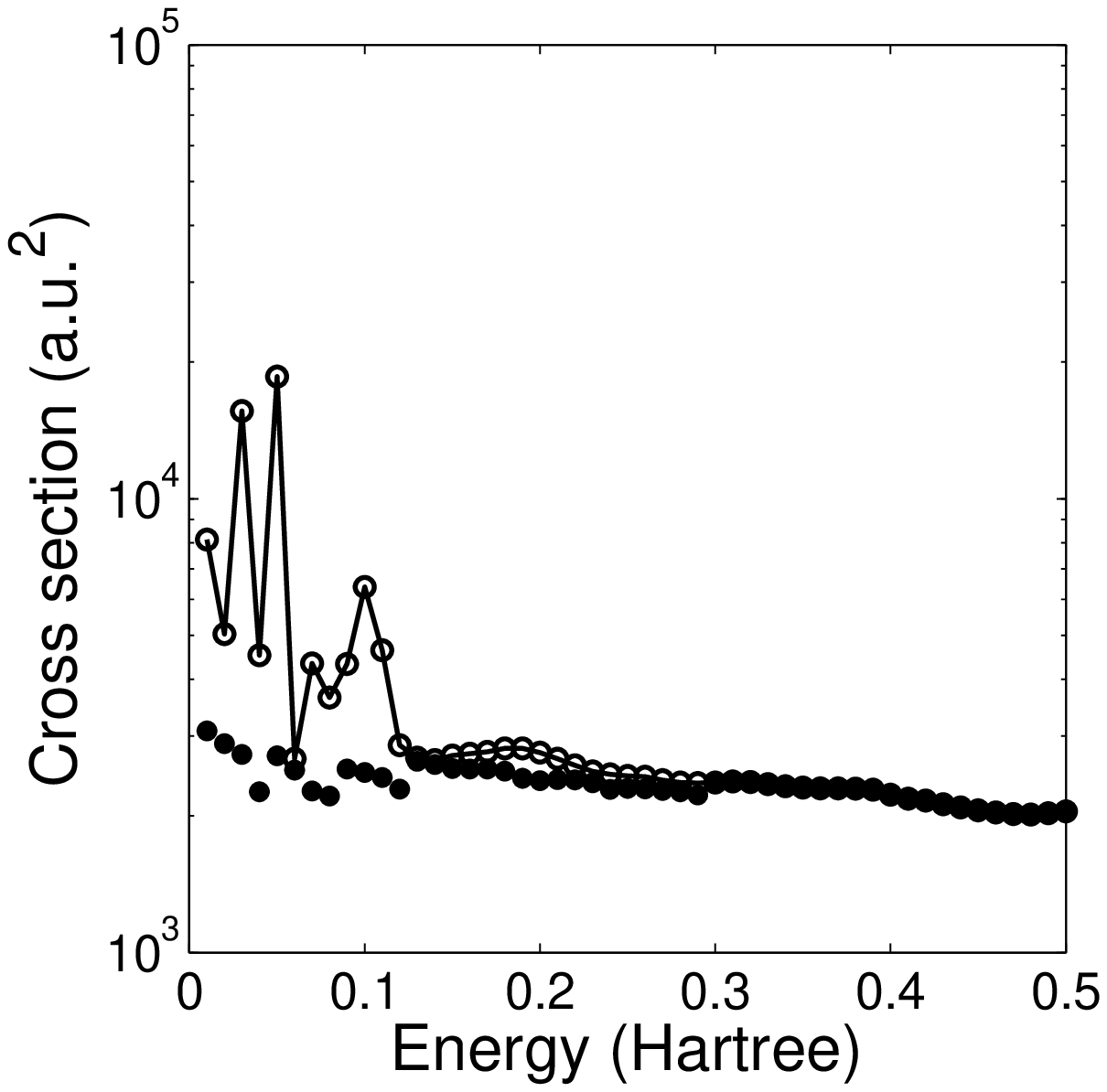}\caption{Total elastic cross
section of the $ \text{poly(A)} \cdot \text{poly(T)}$ decamer as a function of incident electron energy
using the full angular momentum content of the scattering matrix (full line) or 
imposing a strict application of the angular momentum cutoff criterion (dark circles).}%
\label{fig_sigma8andInt_AT}%
\end{figure}

Let us now naively use these rotated $S$ matrices in Eq. (\ref{eqelastic})
using the full extent of angular momenta $l\leq8$ obtained in the R-matrix
calculations for an incident plane wave having a wave vector $\vec{k}$
perpendicular to the mid-section base-pair direction and the spiral axis. Fig.
\ref{fig_sigma8andInt_AT} shows the total elastic cross section as a function of
energy. One immediately notices the surprisingly large and suspicious values
at the lower energies. The two dominant peaks correspond to some fifteen times
the geometrical cross section of the decamer. This is not a low-energy dipolar
effect as all dipoles are cutoff at 12 a.u. in our calculations. As we will see later on, the
situation is even more extreme for the other decamer. The reason for this
low-energy unruly behavior is made clear when one tries to use R-matrix
results for the H$_{2}$O molecule in solid ice \cite{KKR2007}. The correct
band structure of ice can only be obtained at low energy by cutting off the
angular momenta to $l\leq2$ instead of using the full range $l\leq4$. This is not
fortuitous and it can be understood by the following semi-classical argument. For an electron
with angular momentum $L\sim kr$, one can write $L^{2}\approx l(l+1)=k^{2}r^{2}$. If
$r\gtrsim d_{m}$, where $d_{m}$ is the inter-molecular distance, then obviously
the electron is outside of the interaction space of the two molecules. This means
that the relevant angular momenta are those for which $r\lesssim d_{m}$ \textit{i.e.}
$l(l+1)/k^{2}=l(l+1)/(2E_{e})\lesssim d_{m}^{2}$ where energies are in Hartree. But
what is the mathematical or rather the numerical reason? Let us look
closer at Eq. (\ref{eqelastic}). It turns out that the spherical Hankel
function of the first kind $h_{l_{1}}^{(1)}(kR_{nn^{\prime}})$, where
$R_{nn^{\prime}}$ is the distance between the two scatterers $n$ and
$n^{\prime}$, diverges as $(kR_{nn^{\prime}})^{-(l_{1}+1)}$ for small values
of the argument $(kR_{nn^{\prime}})$. The singularity is dominated by the
nearest neighbor distance $d_{m}$. For $l,l^{\prime}_{2}=4$, $l_{1}$ can be as large as
$2l=8$. The reason for this comes from the
angular momentum composition rule that says the sum over $l_{1}$ can reach the uppermost
value of $l_{1} \le \left\vert l+l^{\prime}_{2}\right\vert$. So even though one
expects the $l^{\prime}_{2}=4$ components of the T-matrix
$T=S-1$ to get smaller as the energy decreases, the product of $h_{l_{1}%
}^{(1)}(kR_{nn^{\prime}})$ with $T$ gets overly large because of the much more
singular behavior of the Hankel function. The situation gets much more acute in
our case since for $l\leq8$ the sum over $l_{1}$ goes up to $l_{1}=16$.

\begin{figure}[tbh]
\includegraphics[scale=.4]{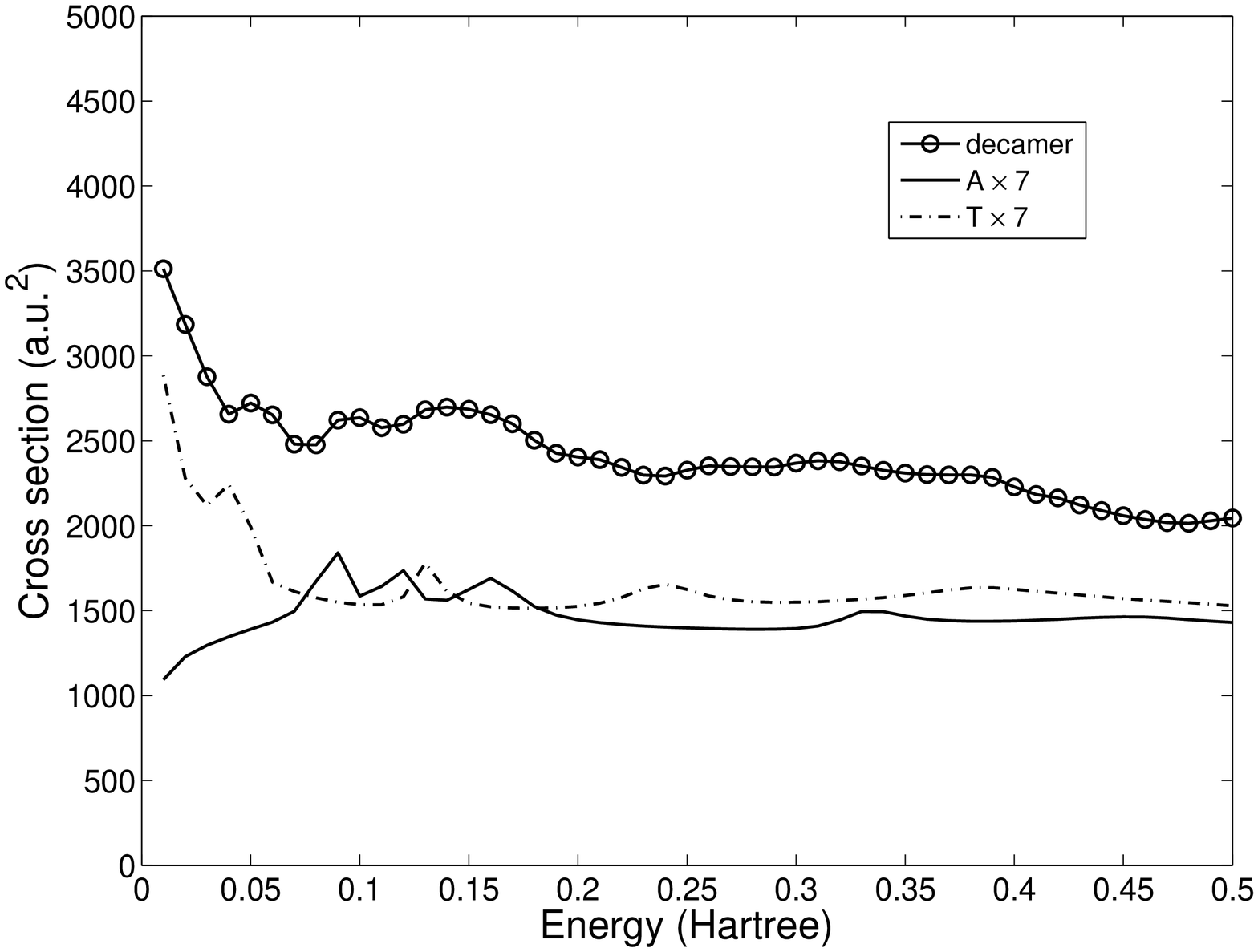}\caption{Interpolated
total elastic cross section of the $ \text{poly(A)} \cdot \text{poly(T)}$ decamer compared with the single
adenine and thymine R-matrix cross section values as a function of incident
electron energy.}%
\label{fig_sigmaInterp_AT}%
\end{figure}

One thus also needs to use a cutoff to describe multiple collisions in the
decamer. Only values $l\leq l_{o}$ should be retained such that $E(l_{o},d_{m})<E_{e}<E(l_{o}+1,d_{m})$
where $E(l_{o},d_{m})=l(l+1)/(2d_{m}^{2})$. One might think that $d_{m}$ should be
equal to the base stacking distance along the spiral axis direction. But it is
clear that an electron that scatters from one molecule to another does so from
every part of the first molecular subunit to every part of the other, including from one
end of the one molecule to the diagonally opposite end of the other one. We have
thus chosen, with some arbitrariness, the value $d_{m}=11$ a.u. which is of the
order of this distance, the size of the bases, the distance between base
centers in the base pairs and
close enough to the size of the R-matrix sphere to retain all of its
important energy dependent characteristics. But this now poses a new problem
having to do with the discreteness of $l$. A strict imposition of these
cutoffs would result in a piecewise chopped cross section as can be seen in
Fig. \ref{fig_sigma8andInt_AT}. One way to circumvent this difficulty, which we have
adopted throughout, is to interpolate any scalar quantity, calculated for all
integer values of angular momenta, at the non-integer value of $l$ obtained
from the solution of $E_{e}=l(l+1)/(2d_{m}^{2})$. A similar procedure is being used
in a study of a H$_{2}$O dimer \cite{Bouchiha2007} and the resulting elastic cross
section is found to be within 5\% of the one calculated using the R-matrix for
energies larger than 2.5 eV. The result of this procedure on the decamer
is shown in Fig. \ref{fig_sigmaInterp_AT}. A comparison with the bare
molecular cross sections obtained from the R-matrix calculation shows good
concordance of most of the peaks, with the exception of the one at 0.24 H in
thymine which is lessened and perhaps split. This is somewhat reassuring for
the credibility of the interpolation procedure.

\begin{figure}[tbh]
\includegraphics[scale=.7]{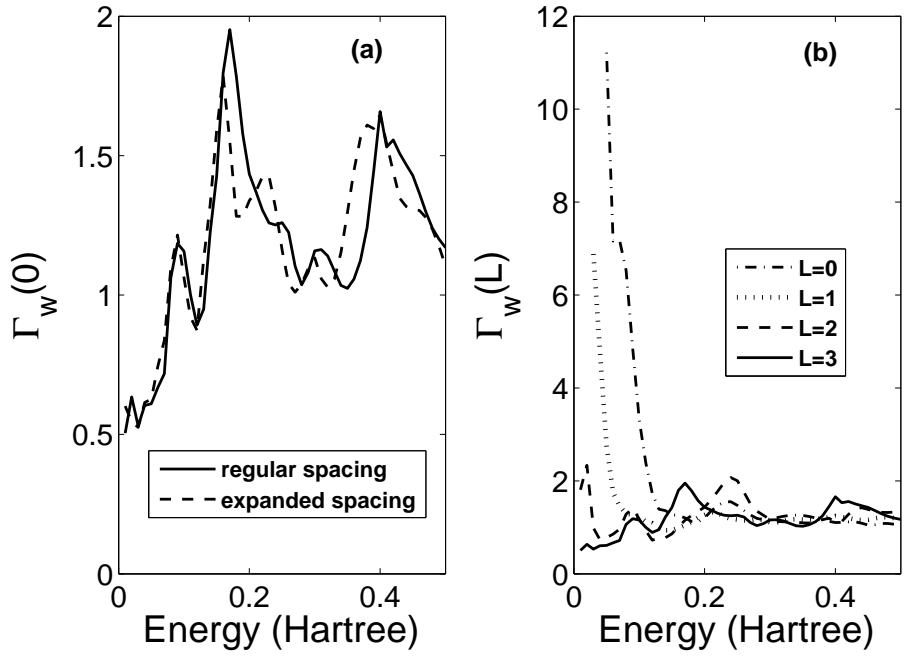}\caption{Interpolated
base averaged value of the square of the electron wave function for the normal
spacing and with an expanded spacing 1.05 times the regular one (a) and interpolated
value of the base averaged weighted partial capture factors (b) for the 
$ \text{poly(A)} \cdot \text{poly(T)}$ decamer as a function of incident electron energy.}%
\label{fig_BO2_compL0andL0to3_AT}%
\end{figure}

\begin{figure}[tbh]
\includegraphics[scale=.4]{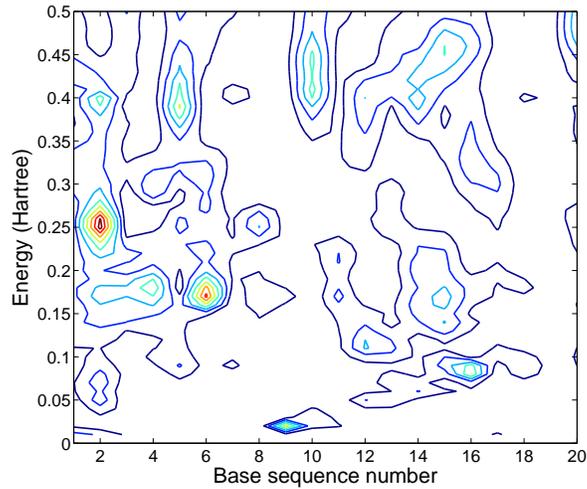}\caption{Contour plot of the
square of the wave function at each base's center for the $ \text{poly(A)} \cdot \text{poly(T)}$ decamer as a function of energy and
base sequence number $n$. Contour lines are equally spaced at integer values.}%
\label{fig_BO2_L0contour_AT}%
\end{figure}

\begin{figure}[tbh]
\includegraphics[scale=.7]{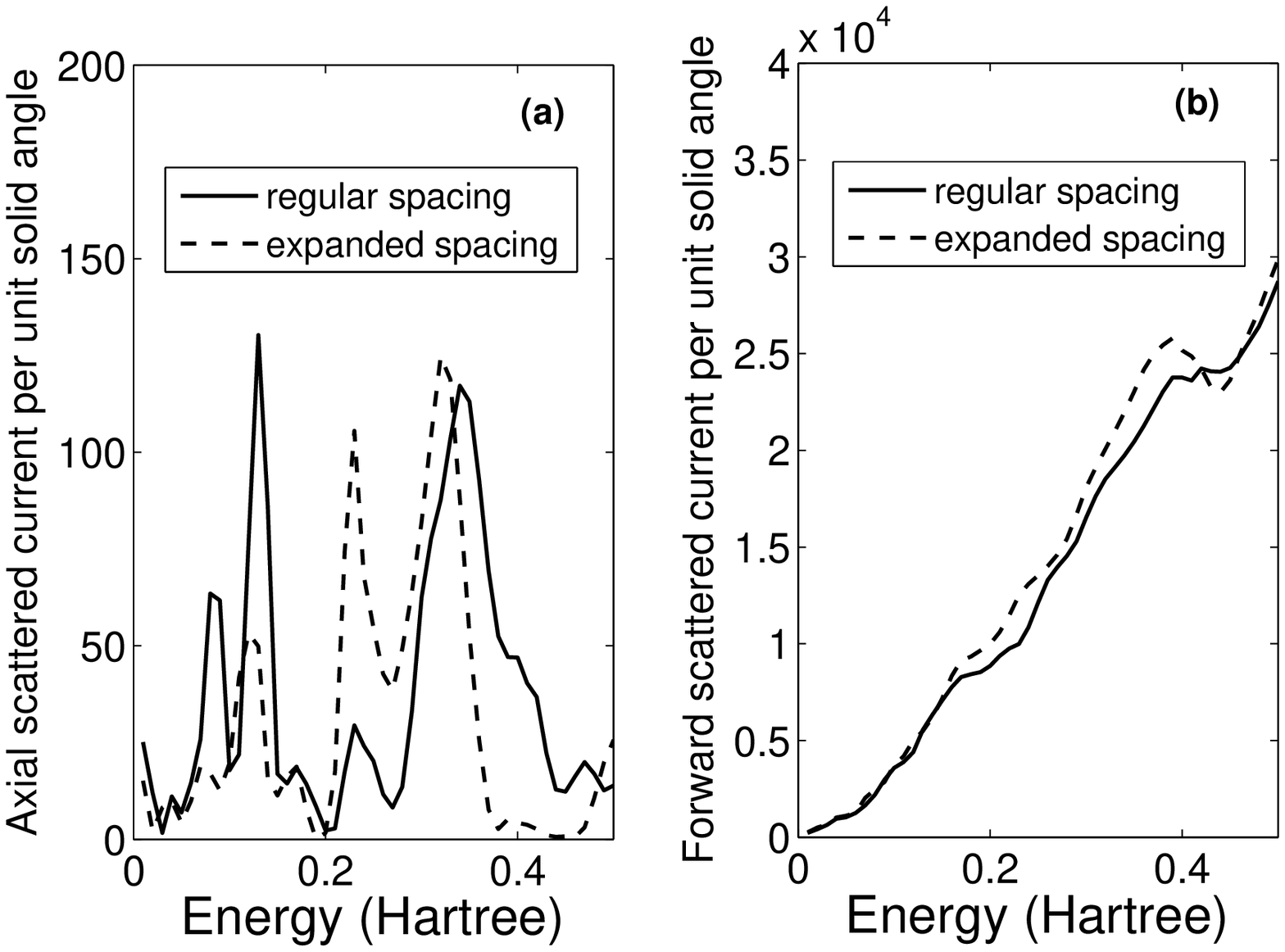}\caption{Interpolated
value of the axial scattered current in the $+\hat{z}$ direction (a) and of the forward scattered current (b)
of the $ \text{poly(A)} \cdot \text{poly(T)}$ decamer as a function of
incident electron energy for the normal spacing and with an expanded spacing
1.05 times the regular one.}%
\label{fig_CAandCF_AT}%
\end{figure}

Let us now look for signs of internal diffraction. As discussed in Refs
\onlinecite{PRL} and \onlinecite{PRA}, this would occur because of the regular spacing of base pairs
along the spiral axis. We have calculated various quantities which were
previously shown to be sensitive to this regular spacing. Fig.
\ref{fig_BO2_compL0andL0to3_AT}(a) shows the quantity $\Gamma_{w}(0)$, equal to the
average of the square of the electron wave function on the bases. It is
compared to the one obtained by dilating the distances along the spiral axis
by a factor 1.05. One expects a down shift in energy for those peaks sensitive
to the inter-base-pair spacing $d_{\parallel}$ by some 10 \% ($E\propto
k^{2}\propto\lambda^{-2}\propto d_{\parallel}^{-2}$). Three peaks seem to
behave this way although the one at 0.42 H at regular spacing is more probably
the internal diffraction one that was deduced in the previous publications.
The unshifted peak slightly below 0.1 H is of local origin.
Fig. \ref{fig_BO2_L0contour_AT} shows a contour plot of $\gamma(0,\vec{R}_{n})$, the absolute square of
the wave function at each base's center, in energy and base sequence space. One must of course realize that 
these contour lines are calculated by interpolating $\gamma(0,\vec{R}_{n})$, which is
defined only for a discrete set of the index $n$. The virtue of such a graph is to
enable a quick survey of cool or hot spots at which the square of the wave function
is small or large.  The sequencing index
varies from 1 to 10 going up the A strand and from 11 to 20 going down the T
strand. The peak structures in $\Gamma_{w}(0)$ showing appreciable enhancement
of the wave function are seen to occur mostly on the base pairs that lie
nearly perpendicular to the incident electron direction (actually, the closest
to perpendicular is at an angle of $90\pm18$ degrees), at the beginning,
middle, and end of the decamer, where phase coherence is more favorable. We
have also calculated the axially scattered current per unit solid angle in the $+\hat{z}$ direction
Fig. \ref{fig_CAandCF_AT}(a) and the forward scattered current per unit solid
angle in Fig. \ref{fig_CAandCF_AT}(b) under identical axial spacing conditions as in Fig. \ref{fig_BO2_compL0andL0to3_AT}(a).
The only shifting peak that is common to all figures is the one at 0.42 H at
regular spacing. This is clearly the internal diffraction peak. One should
note the large peak at 0.34 H in Fig. \ref{fig_CAandCF_AT}(a). There is nothing
special that can be seen in the wave function in Fig.
\ref{fig_BO2_L0contour_AT} at that energy. This result can only be understood
by interference between the scattered beams emanating from the bases. For the
current along the axis, aside from the contribution of the beam amplitudes at
each base $\vec{R}_{n}$, there is an extra exit phase factor at position
$z\rightarrow\infty$ on the axis outside the decamer, which is proportional to
$\exp[ik(z-z_{n})]$. This means that there is an optimal value of $k$ at which
$\exp[ik(z_{n^{\prime}}-z_{n})]$ between neighboring base pairs along the
spiral axis will \textquotedblleft synchronize\textquotedblright\ the beams
and produce an overall good phase coherence. The position of the peak yields
$k\sim0.825$ and a $\lambda\sim7.6$. The vertical distance between base pairs
is $6.4$ a.u.. Why don't the numbers match? This is because we are dealing with a
complex superposition of many different partial waves with different phases
for all $l$ (and $m$) less or equal to 8. The optimal value of $\lambda$ has
to compromise with all of these and this happens for $\lambda\sim7.6$.

We can now also answer the question whether or not there is appreciable
enhancement of the partial capture factors at low energy. Fig.
\ref{fig_BO2_compL0andL0to3_AT}(b) clearly says so for energies less than say 0.1 H, a
region of interest for the low-energy $l=2,3$ shape resonances of the bases
(see Fig \ref{fig_sigmaInterp_1000and20}(a) showing cross sections of the 4 bases).

\subsection{B-form GCGAATTGGC base pairs decamer}

\begin{figure}[tbh]
\includegraphics[scale=.7]{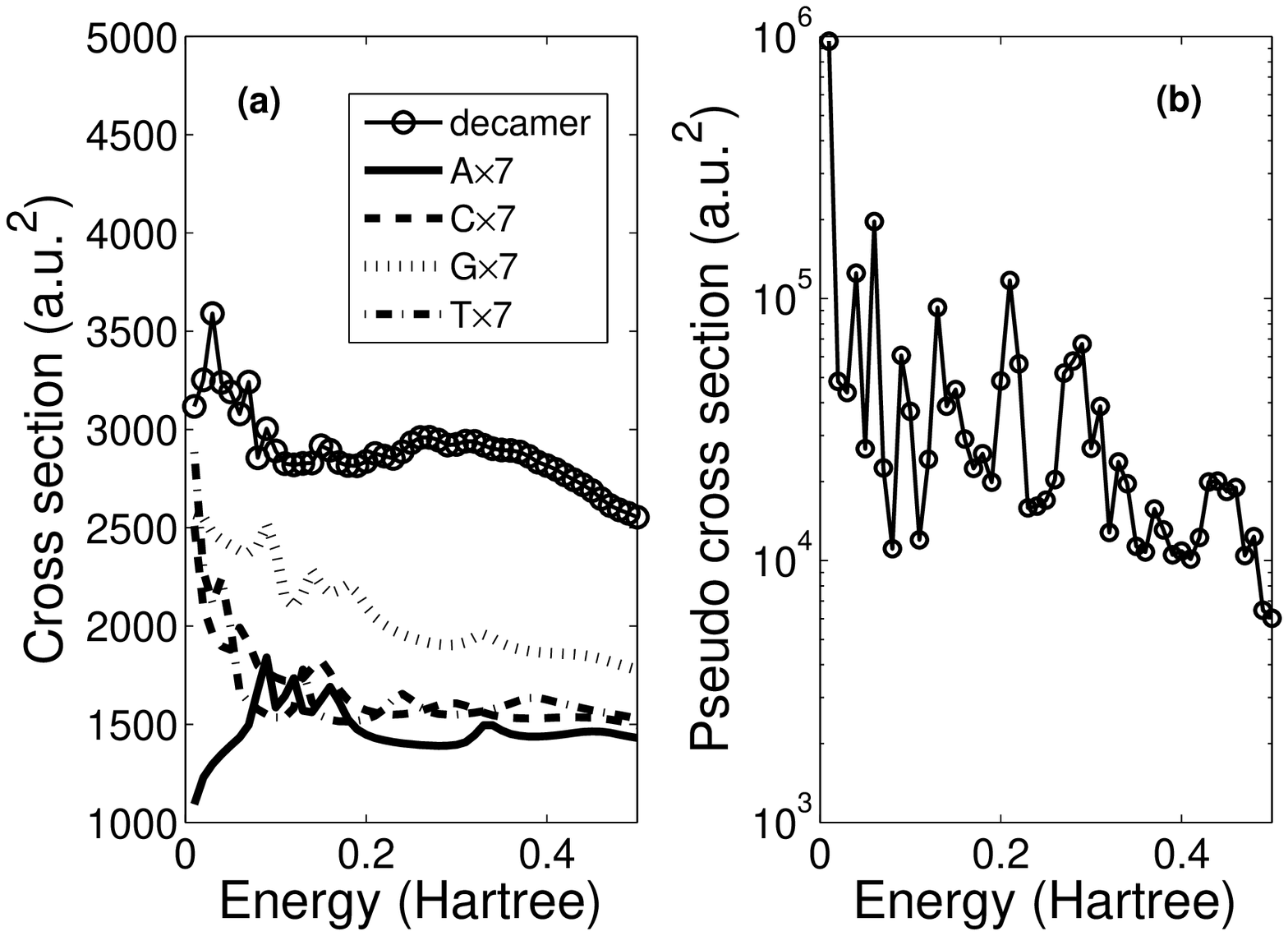}\caption{Interpolated
total elastic cross section of the GCGAATTGGC B-form decamer at $\xi=1000$ 
compared with the single adenine, cytosine, guanine, and 
thymine R-matrix cross section values (a) 
and at $\xi=20$ (b) as a function of incident electron energy.}%
\label{fig_sigmaInterp_1000and20}%
\end{figure}

\begin{figure}[tbh]
\includegraphics[scale=.5]{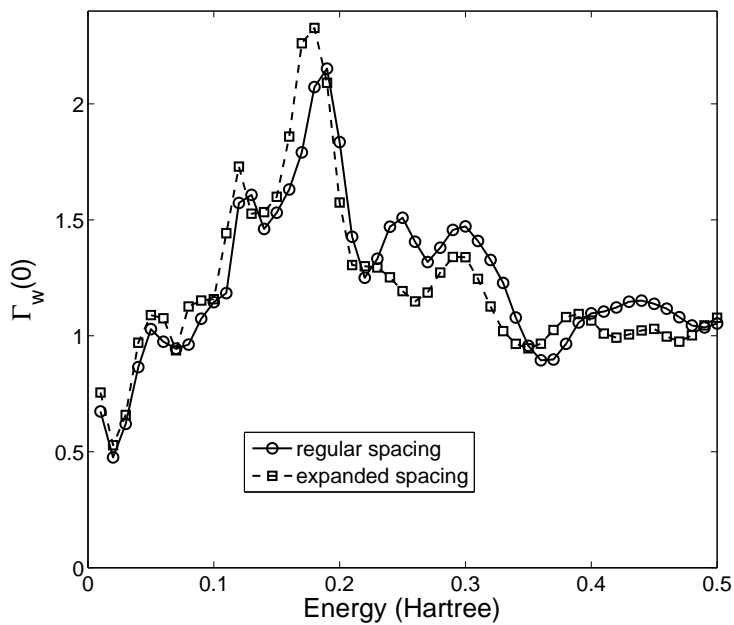}\caption{Interpolated
value of the average of the square of the electron wave function on the bases
of the GCGAATTGGC B-form decamer at $\xi=1000$ as a function of incident electron
energy for the normal spacing and with an expanded spacing 1.05 times the
regular one.}%
\label{fig_BO2_compL0_1000}%
\end{figure}

\begin{figure}[tbh]
\includegraphics[scale=.4]{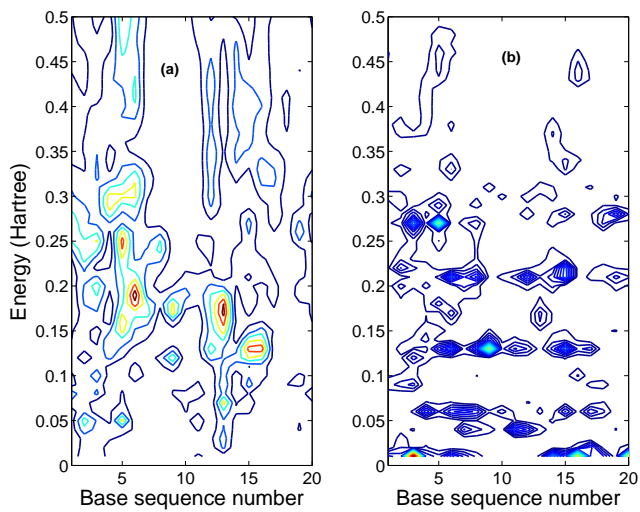}\caption{Contour plot of the
square of the wave function at each base's center for the GCGAATTGGC B-form decamer at $\xi=1000$ (a) and
at $\xi=20$ (b) as a function of energy and base sequence number $n$.
Contour lines are equally spaced at integer values in (a) and multiples of 5 in (b)}%
\label{fig_BO2_L0contour_1000and20}%
\end{figure}

\begin{figure}[tbh]
\includegraphics[scale=.7]{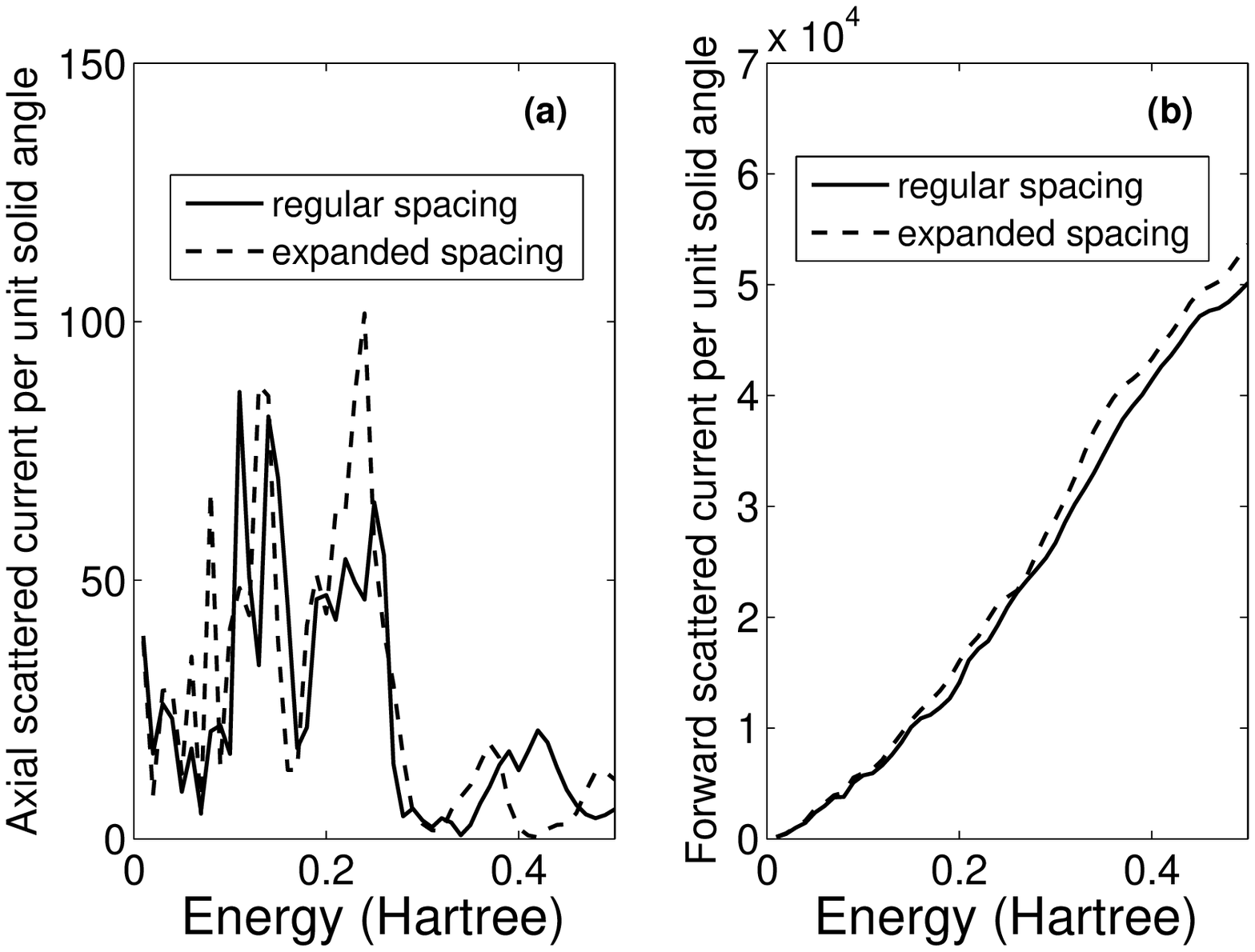}\caption{Interpolated
value of the axial scattered current in the $+\hat{z}$ direction (a) and the forward scattered 
current (b) of the GCGAATTGGC B-form decamer at $\xi=1000$
as a function of incident electron energy for the normal spacing and with an
expanded spacing 1.05 times the regular one.}%
\label{fig_CAandCF_1000}%
\end{figure}

Now that we know how to look for internal diffraction, let us repeat the
analysis of the last section on our sequence disordered decamer for $\xi=1000$
a.u. and $\vec{k}$ perpendicular to the mid-section base-pair direction and
the spiral axis. The total elastic cross section as a function of energy using 
the full extent of angular momenta $l\leq8$ exhibits a cross section that has 
peaks some 500 times the geometrical one. This confirms the unphysical 
behavior of this procedure at low energy.
Using the interpolation procedure, however, restores a credible cross section
as shown in Fig. \ref{fig_sigmaInterp_1000and20}(a). A comparison with the cross
section of individual bases again reveals a satisfactory correspondence of
most peak structures. Fig. \ref{fig_BO2_compL0_1000} shows the quantity
$\Gamma_{w}(0)$, equal to the average of the square of the electron wave
function on the bases. It is compared to the one obtained by dilating the
distances along the spiral axis by a factor 1.05. The diffraction signature
around 0.4 H is still present but with an appreciably reduced enhancement
factor relative to the plane wave value of 1. This is the effect of sequence
disorder which turns out to be stronger than what was anticipated from the toy
model calculations. The lower energy structures are similar to those of the
$ \text{poly(A)} \cdot \text{poly(T)}$ decamer. Fig. \ref{fig_BO2_L0contour_1000and20}(a) 
shows a contour plot of $\gamma(0,\vec{R}_{n})$, the absolute square of
the wave function at each base's center, in energy and base sequence space. The
sequencing index varies from 1 to 10 going up the GCGAATTGGC strand and from
11 to 20 going down the complementary strand. The peak structures in
$\Gamma_{w}(0)$ of Fig. \ref{fig_BO2_compL0_1000} showing appreciable enhancement of the wave function are seen
to occur mostly in the mid region, on the base pairs that lie mostly
perpendicular to the incident electron direction. Sequence disorder has a
visible effect on the interference patterns within the decamer and not only on
internal diffraction. It somewhat desynchronizes the ends from the
mid-section. The somewhat irregular distribution of the peaks at lower
energies ($E\lesssim0.3$ H) in both decamers studied thus far shows that these
structures are local ones and perhaps a signature of weak localization.
Localization has been predicted in  $ \text{poly(dA)} \cdot \text{poly(dT)}$ decamer studies due to
structural changes promoted by thermal fluctuations \cite{Lewis2003}. There is
a parallel to be made between the time evolution of localized states and our
energy dependence. The axially scattered current in the $+\hat{z}$ direction (from the G end to the C end) shown in Fig. \ref{fig_CAandCF_1000}(a)%
\ also presents a clear but quite diminished internal diffraction peak. But the
large peak at 0.34 H in Fig. \ref{fig_CAandCF_AT}(a) has now disappeared,
stressing the importance of sequence disorder. The forward scattered current
of Fig. \ref{fig_CAandCF_1000}(b) shows no evidence of internal diffraction.
This parameter is less sensitive to internal diffraction as was already
evident from the $\text{poly(A)}\cdot \text{poly(T)}$ results.  The weighted 
partial capture factors are very similar to the the $\text{poly(A)}\cdot \text{poly(T)}$ one.

Let us now reduce the coherence length to $\xi=20$ a.u. to simulate phase decoherence
within the decamer. The pseudo cross section $\sigma_{pe}$ of Eq. (\ref{eq_pseudosigma}%
) is shown in Fig. \ref{fig_sigmaInterp_1000and20}(b).   It is very jagged compared to
the previous cross sections. Remember that the values are deceptive since
$\sigma_{e}$ decreases by a factor of fifty at a distance of 40 a.u. from the
geometrical center of the decamer and only a few a.u. from the ends. A similar 
behavior for $\Gamma_{w}(L)$ can be understood in terms of huge localization 
resonances very apparent in Fig. \ref{fig_BO2_L0contour_1000and20}(b), which shows the contour plot 
of $\gamma(0,\vec{R}_{n})$, the absolute square of
the wave function at each base's center, in energy and base sequence space. Such localization resonances were seen in highly
disordered single-strands of DNA \cite{PRA3} although not as pronounced. 
There is localization at all energies although it gets weaker as the energy
increases. There is no obvious pattern in their position within the decamer.
The enhancement of the partial capture factors is very large in all resonances
with energy less than 0.3 H. These large numbers are an artifact
of a large imaginary part to the wave number which prevents bases from ``seeing" their far
neighbors and thus forces a network of strong correlation of amplitudes with the near neighbors. 
This is understandable in the presence of important inelastic losses. This appears 
inappropriate, however, for disorder due to stray elastic scatterers since
correlations will still exist with all bases even though there is randomization of
the local phases. These correlations should weaken this network. It would be important to verify this
by adding some genuine randomness such as including water molecules bound to DNA. We
plan to do this in the near future.

\subsection{A-form GCGAATTGGC base pairs decamer}

We finally look at the A-form of the GCGAATTGGC base pairs decamer. There is considerable 
tilt (22.6 degree inclination) in addition to roll (-10.5 degrees propeller twist) of 
the base pairs in this form \cite{Neidle}. Moreover, the pairs are more tightly packed 
having a rise of 2.54 \AA\ compared to 3.38 \AA\ for the B-form and there are 11 base
pairs per turn of the helix. It is of interest to look for the differences in the scattering results.

\begin{figure}[tbh]
\includegraphics[scale=.4]{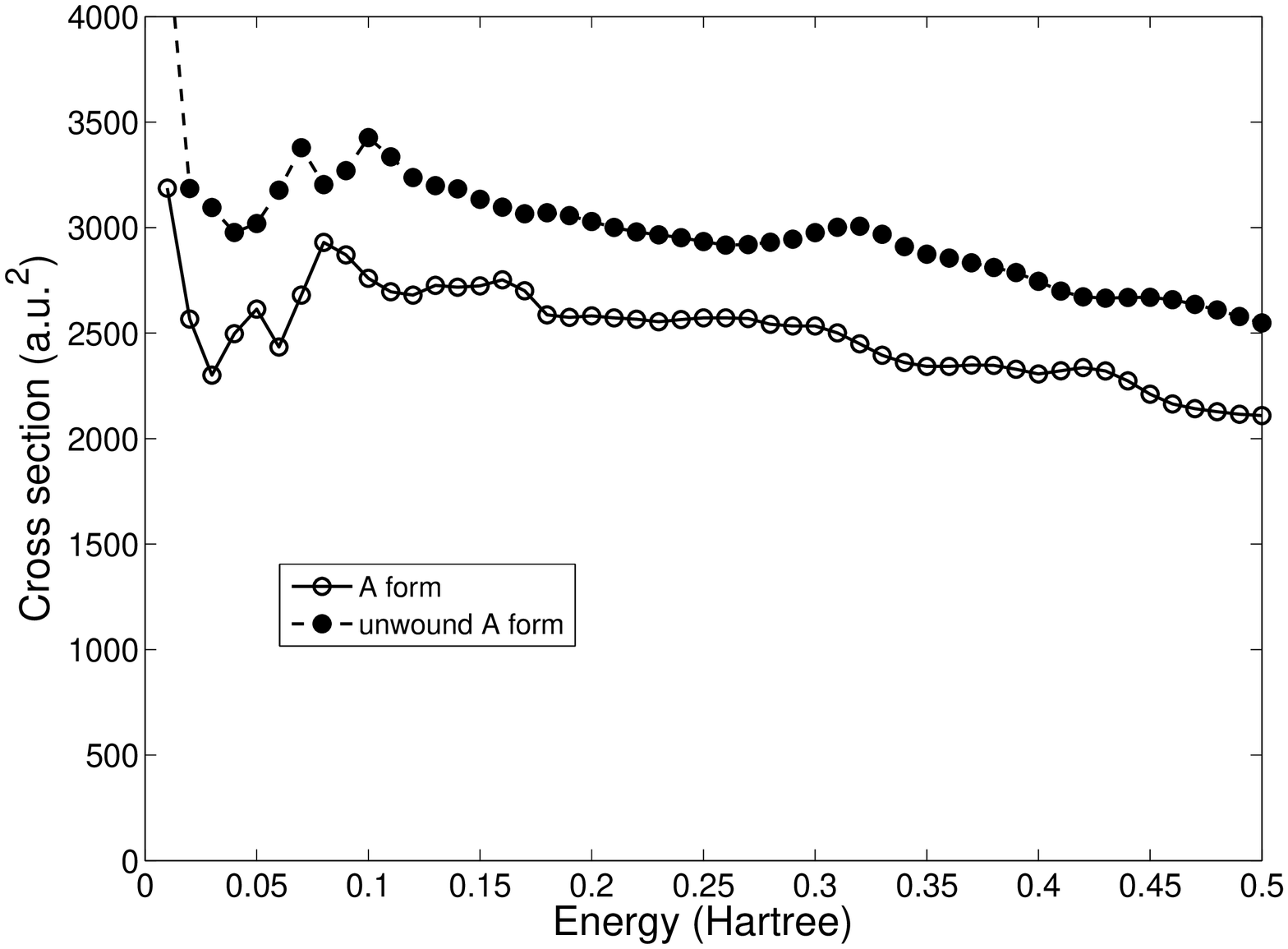}\caption{Interpolated
total elastic cross section of the GCGAATTGGC A-form decamer at $\xi=1000$ (open circles) 
and of the unwound counterpart having B-form rise and helical twist (black circles) as a function 
of incident electron energy.}%
\label{fig_CS_AformWithTRtoB}%
\end{figure}

\begin{figure}[tbh]
\includegraphics[scale=.4]{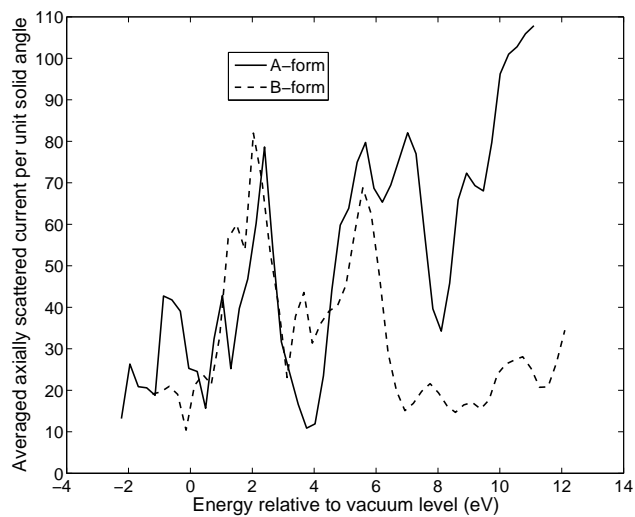}\caption{Average of the
scattered current in the two axial directions and for two incident and mutually 
orthogonal directions of the GCGAATTGGC A-form and B-form decamers at 
$\xi=1000$ as a function of incident electron energy.}%
\label{fig_axial_AB}%
\end{figure}

\begin{figure}[tbh]
\includegraphics[scale=.4]{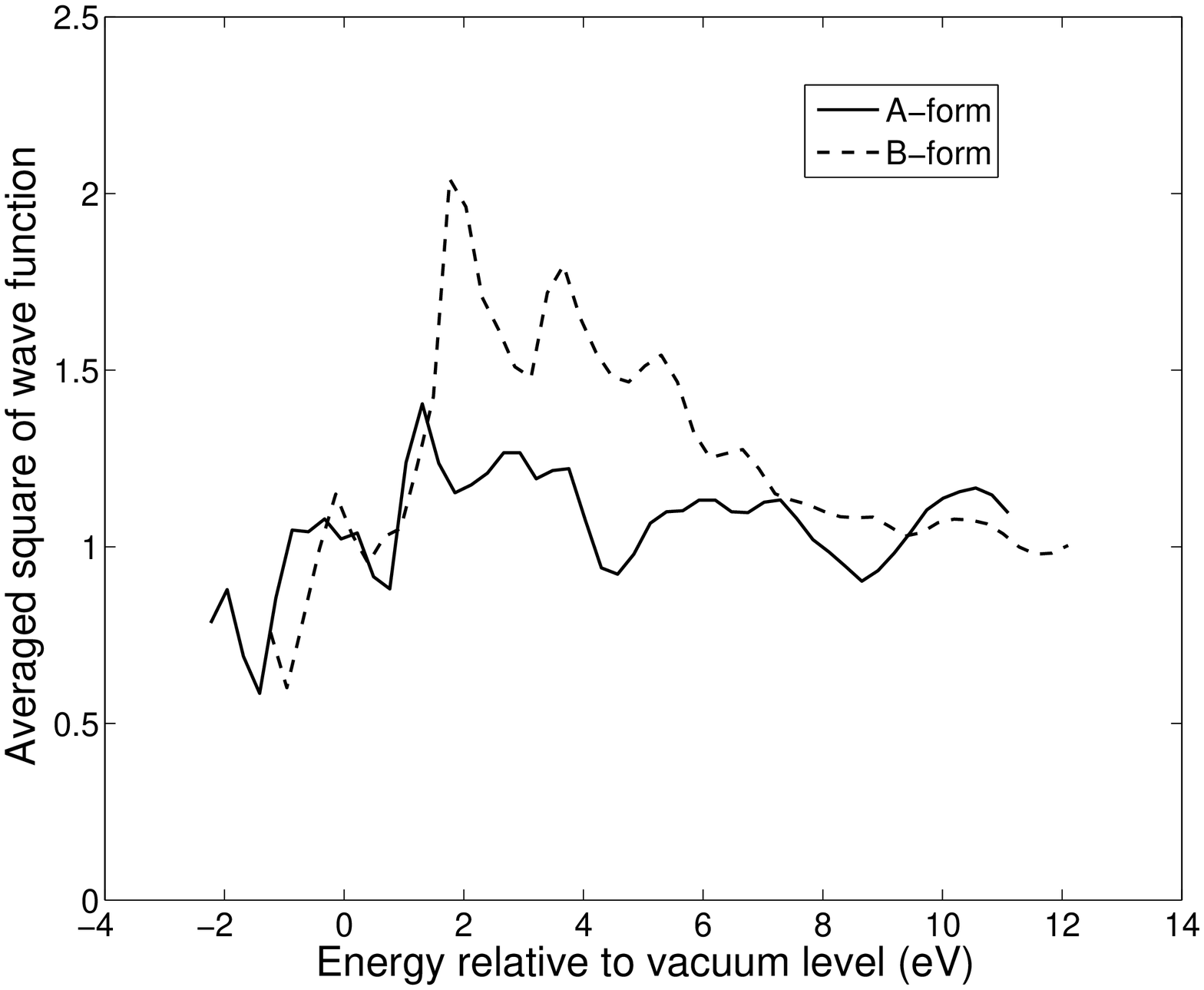}\caption{Average of the square of the 
electron wave function on the bases over two incident and mutually 
orthogonal directions of the GCGAATTGGC A-form and B-form decamers at 
$\xi=1000$ as a function of incident electron energy.}
\label{fig_B02_AB}%
\end{figure}

Fig. \ref{fig_CS_AformWithTRtoB} shows the cross section for this A-form decamer. One immediately
notices the globally smaller values compared to its B-form analog in Fig. \ref{fig_sigmaInterp_1000and20}(a).
This was unexpected in view of the closer packing of the base pairs and of the results from the toy 
model \cite{PRA} which led us to expect larger capture factors. This emphasizes 
the need for realistic scattering models. It would seem that there 
is some increasing destructive interference as the base pairs get closer.
In order to validate this hypothesis, we have unwound the A-form by increasing the rise and 
helical twist to agree with the B-form values, but keeping the base pair tilt and roll intact.
Fig. \ref{fig_CS_AformWithTRtoB} shows the cross section for this unwound decamer. 
One recovers the same scale of values as for the B-form decamer. It is therefore indeed destructive interference 
caused by the closer proximity of the base pairs that is responsible for the smaller cross section
of the A-form and not the tilt and roll. The latter are, however, surely responsible for the
difference in peak structures. Note that there is no internal diffraction to be seen in the A-form in the chosen energy range. 
The distance between nearest base pairs (the rise) is too small which pushes the diffraction effects to energies beyond 0.5 H.

Finally, we have computed scattering parameters for both the A and B forms which are 
more easily comparable with existing experimental data.  The averaged electron current 
scattered along the DNA axis ($\pm$Z scattered directions and X, Y incident directions) 
is shown in Fig. \ref{fig_axial_AB}, as a function of electron energy. 
Similarly, the magnitude of the square of the wavefunction averaged over 
all bases is shown in Fig. \ref{fig_B02_AB}. To facilitate comparison with experimental data, the energy scale has 
been converted to eV and the zero electron energy shifted by the estimated 
value $U_{op}$ of the polarization energy. This shift is necessary, since in thin film 
experiments with electrons incoming from vacuum, the zero energy is conveniently 
referenced to that of the vacuum level. The $U_{op}$ values for the A and B forms were 
evaluated by estimating the polarization energy on a base due to its near neighbors. As we 
wish to make comparisons with experimental results on DNA, we have added an interaction with the nearest 
phosphate and deoxyribose subunits as well as to some four \cite{structural_data} close structural water molecules that would be present in real DNA.
We have utilized the formula $U_{op}\approx -\sum \alpha/(2d^4)$ where 
$\alpha$ was given an average value of 100 $(a.u.)^3$ for the in-plane polarizability of the partner base \cite{Basch:CPL1989}, 
an average value of 47 $(a.u.)^3$ for the out of plane polarizability of the two axial neighbors \cite{Basch:CPL1989},
an average value of 5 and 80 $(a.u.)^3$ for the the isotropic polarizabilities of phosphate and deoxyribose respectively \cite{ChemSketch}, and 
of 10 $(a.u.)^3$ for the water molecules \cite{Olney1997}. We used a distance $d\sim 11$ a.u. for the 
partner distance and 7 a.u. for all other distances \cite{structural_data} except for the A-form axial nearest 
neighbor distance, which was taken to be 5.3 a.u.. Needless to say that the values obtained for 
$U_{op}$, -1.5 eV and -2.5 eV for the B-form and A-form respectively, are approximate although credible.

\section{Comparison between theory and experiments}

Low-energy electron (LEE) experiments with DNA have been performed on dry films, in 
ultra-high vacuum, where the molecule adopts the A-form owing to non-structural water 
loss \cite{Swarts92}.  These experiments have measured the damage inflicted to DNA 
mostly by electrons in the 0-20 eV range, i.e., the yield functions, in the form of 
base \cite{sanche2003,sanche4}, sugar \cite{Cai:JCP2006} and phosphate lesions 
\cite{Pan2005}, single and double strand breaks \cite{Sanche:DNA,Sanche:RPD2002,sanche2003a} 
and base release \cite{Zheng2005}.  Thus, no electron scattering experiments are presently 
available to be directly compared to our theoretical results.   However, some of the 
mechanisms, which have been invoked to account for the magnitude of the yield of specific 
damages and peak energies in their yield functions below 15 eV, rely on scattering 
properties of LEE within the DNA molecule.

In experiments with thin films of plasmid DNA the yield function for single strand 
breaks (ssb) exhibits maxima at 0.8, 2.2 and 10 eV, with a shoulder at 6 eV 
\cite{Sanche:DNA,Sanche:RPD2002,Sanche_Burrow:PRL04} which appears as a distinctive 
peak at this energy in the ssb yield function of synthetic single stranded DNA 
films \cite{Sanche_bond:JCP06}.  The peak positions were determined with an accuracy 
of ± 0.3 eV. From the analysis of LEE-induced products from this latter type of films, 
the yield function for base release was also found to exhibit peaks at 6 and 10 eV 
\cite{Sanche_bond:JCP06}.  Below 15 eV, the yield function for the induction of 
double strand breaks (dsb) was found to be dominated by a single peak located 
at 10 eV \cite{Sanche:DNA,sanche2003a}.  Owing to their low energies the 0.8 and 
2.2 eV maxima could easily be interpreted as shape resonances \cite{Sanche_Burrow:PRL04}, 
but the exact mechanism leading to rupture of the C-O phosphodiester bond (i.e., ssb) 
\cite{Zheng2005} is not obvious for two reasons.  First, the cross section for 
ssb induction below 3 eV is of the order of 10-17 cm2 per base pair \cite{Panajotovic:RR2006}, 
which is fairly large for damage caused by a single anion dissociative state at 0.8 
and 2.2 eV, respectively.  In fact, this value is almost as large as that measured 
at 100 eV, in the same DNA, where a plethora of ionization and dissociation channels 
are available \cite{sanche2003a,Panajotovic:RR2006}.  Coherent enhancement of the 
electron wavefunction within DNA may therefore increase at low energies the electron 
capture cross section of transient anion states responsible for ssb.  Secondly, in 
the experiment of Martin et al. \cite{Sanche_Burrow:PRL04}, the 0.8 and 2.2 eV peaks 
did not coincide with the energy of the dissociative phosphate anion which is known to 
rupture the C-O bond of the DNA backbone \cite{Zheng2005}. Instead, these peaks coincided 
with those in the electron capture cross section of the DNA bases. The latter result 
lead Martin et al. \cite{Sanche_Burrow:PRL04} to postulate that the electron was first 
captured by the basis and then transferred to the phosphate group.

In fact, according to the theoretical studies of Simons' group, below 3 eV electrons cleave 
the CO bond of the DNA backbone at the 3' and 5' positions essentially via electron transfer 
\cite{Barrios:JPCB02,Berdys2004,Berdys:JPCB2004,Berdys:JPCA2004}.  An electron is first captured 
by a base to form a pi* transient anion and afterwards the additional electron transfers, 
via the sugar moiety, to the phosphate unit where it occupies a sigma* orbital at either the 3' 
or 5' C-O positions.  The resulting anion state being dissociative leads to C-O bond cleavage 
(i.e., a ssb). Electron transfer is not necessarily limited to the nucleotide where capture 
occurs \cite{Wagenknecht:NPR2006,Giese:Nature2001}. Thus, if the main mechanism leading to 
ssb is electron transfer from the bases, the large cross section for damage below 3 eV 
could be explained by invoking strong constructive interference of the electron wave function 
due to stacking of the bases along the DNA chain.

This cross section could maximize at 0.8 and 2.2 eV, if the incoming electron preferably 
scatters along the DNA axis at these energies, where it would cause preferential coherent 
enhancement due to base stacking. This is exactly what is found in the energy dependence 
of the magnitude of the electron current scattered along the axis of A-DNA in Fig. 
\ref{fig_axial_AB}: the current in the Z-direction is maximized at 0.8 and 2.2 eV.  When 
DNA is modified to its B form the two peaks move closer to each other and little correlation 
is found with the experimental values.  In Fig. \ref{fig_B02_AB}, the average of the 
square of the scattered electron wavefunction on a base is found to maximize at 1.3 and 
2.8 eV in fair agreement with the experimental maxima in the yield function for ssb. 
For the B form, Fig. \ref{fig_B02_AB} shows no two-peak correlation between theory and 
experiment.  Thus, by including both the shape resonance wavefunctions and constructive 
interference due to base stacking our calculation can represent fairly well the energy 
dependence of the yield of ssb below 3 eV.

The broad 6-eV feature in the experimental ssb yield function of DNA, which spreads 
from 5 to 7 eV \cite{Sanche_bond:JCP06}, is also in good agreement with the theoretical 
results of Figs \ref{fig_axial_AB} and \ref{fig_B02_AB} for the A-form.  Both curves exhibit two peaks 
around 6 eV, which if unresolved would produce a broad maximum around 6 eV, as observed 
experimentally.

The 6-eV feature has been studied in detail by Zheng \textit{et al.} who bombarded thin molecular 
films of a short single strand of DNA, with electrons of energies between 4 and 15 eV 
\cite{Sanche_bond:JCP06}.  By high-pressure liquid chromatography, they identified 12 
fragments of the oligonucleotide GCAT sequentially composed of the bases guanine (G), 
cytosine (C), adenine (A) and thymine (T).  The yield functions exhibited maxima at 6 
and 10-12 eV, which were interpreted as due to the formation of transient anions leading 
to fragmentation. Later, they analyzed the products induced by 4-15 eV electrons incident 
on two abasic forms of the tetramer GCAT, i.e., XCAT and GCXT, where X represents the 
base, which has been removed and replaced by a hydrogen atom \cite{Zheng:PRL2006}.  
The results obtained at an incident energy of 6 eV showed that essentially no strand 
break occurs at positions in the backbone corresponding to those of the missing base.  
\textit{This finding clearly indicated that at 6 eV, and possibly below, electrons break 
the DNA backbone almost exclusively via electron transfer}.  Furthermore, the total yield 
of all the bases released and ssb induced by electrons were found to be strongly affected 
by the presence of an abasic site; in both XCAT and GCXT, the yield of detached bases was 
found to be up to an order of magnitude smaller than that from GCAT \cite{Zheng:PRL2006}. 
Thus, the initial electron capture amplitude was suggested to be highly sensitive to the 
number and possibly the geometrical arrangement of the bases, indicating the presence of 
a strong collective effect.  According to the results of Figs \ref{fig_axial_AB} and 
\ref{fig_B02_AB}, this collective effect could be related to strong electron scattering 
on axis around 6 eV, which exhibits an 8-fold increase in magnitude in going from 4 to 6 
eV. Coherence enhancement of the scattered electron wavefunction around 6 eV could also 
play an important role.  Both phenomena should be highly sensitive to base removal, since 
they are directly related to base stacking and their periodicity.

Finally, the maximum at 10 eV in the electron energy dependence of the calculated averaged 
axial scattered current and square of the wavefunction averaged on all bases for the A-form 
correlate well with the strong maximum found experimentally in the yield function for ssb 
and dsb induced by LEE impact on plasmid DNA \cite{Sanche:DNA,sanche2003a}.  This result 
suggests that coherence also influences DNA damage in the 10 eV region; i.e., at energies 
where the formation of core excited resonances is the dominant mechanism implicated in bond 
rupture \cite{Sanche:EurPhys2005}. Such anion states, which consist of two electrons 
occupying electronically excited orbitals around a positive hole of a DNA subunit, are highly 
localized \cite{Sanche:EurPhys2005}. It is not obvious how such anion states would couple to 
a diffracted wave along the DNA axis, a problem which has not been addressed in the present 
work. Furthermore, present and previous \cite{PRA,PRA2} partial wave analysis of the enhanced 
electron capture probability on specific subunits due to diffraction has been show to be much 
reduced in the 10 eV region compared to low energies. It is therefore possible that the energy 
coincidence mentioned above is fortuitous. 

To check the validity of all of our comparisons, we calculated the averaged electron current 
scattered along the DNA axis (Z direction) and the magnitude of the square of the wavefunction 
averaged over all bases for different base sequences of the A-form. We observe a fairly good 
stability in the features of the curves up to about 7 eV. In other words, the effect of 
diffraction is not very dependent on sequence and therefore a comparison with experimental 
data obtained with different sequences can be considered significant below 7 eV.  Thus, the 
stability of the structural information in Figs \ref{fig_axial_AB} and \ref{fig_B02_AB} with 
different base sequences and the comparison with experimental results indicates that wave function interference 
should be taken into account to describe the mechanism of action of electrons with energies 
lower than 7 eV in DNA. Beyond this value, the energy of the calculated minima and maxima change 
according to sequence. Thus in this case, multiple scattering of the electron wave in DNA is 
highly sequence dependent and comparison with data obtained with plasmid DNA, which is longer 
and has a different sequence, is not considered to be significant.

\section{Conclusions}
This article describes the first attempt to use high-quality gas phase
scattering data to gain insight into the interaction, in condensed phase, of
DNA and low energy electrons, which is relevant for radiation damage to
nucleic acids by ionizing radiation. In particular, we used R-matrix
calculations performed by some of us \cite{Tonzani:JCP05} for the gas phase scattering. 
To explore the interaction of a continuum electron with the DNA double strand,
we couple the
gas phase calculations with a multiple scattering
framework developed by some of us \cite{PRL,PRA,PRA2,PRA3} a few years ago. The
combination of R-matrix calculations and multiple scattering has been recently
shown \cite{KKR2007,Bouchiha2007} to give accurate results for simpler systems as the water dimer and ice.
These recent works have also inspired the angular momentum cutoff procedure and
the interpolation of the scalar quantities described in Sec. \ref{sec:results}.

The results we show here, in a similar fashion to Refs.
\onlinecite{PRL,PRA,PRA2,PRA3}, show that some peaks in electron scattering from
macromolecules can be due to diffraction instead of formation of negative
anions. They also show that single basis calculations are not enough to gain an
understanding of the process and that collective effects can be as
important as local ones.
Predictably, disorder is shown to play an important role, as evidenced from the
effect of sequence disorder and of electron decoherence that we
have performed introducing an imaginary part of the optical potential, as in
Sec. \ref{sec:mult_scat}. It is difficult to give a quantitative assessment of
the role of disorder in electron scattering from biomolecules within our
current model, but it would be definitely an interesting topic unto itself,
also for investigations by the experimental community, maybe performing
measurements in solution at different temperatures.  
We have shown how decoherence leads to strong resonance localization and large wave function amplitudes,
generating strong resonance enhancements across the energy range we explored.

Despite the simplicity of our model, we found fairly good  correlation between
calculated values for the A form and experimental results obtained on
LEE-induced damage to DNA in this configuration.  Below 7 eV, the calculated
averaged electron current scattered along the DNA axis (Z direction) and the
magnitude of the square of the electron wavefunction, averaged over all bases
for different base sequences of the A-form, were found to be essentially
independent of the nature and sequence of the bases.  It was therefore possible
to compare calculated values with the results of experiments performed with a
longer DNA molecule of a different sequence.  
The structures which appeared in the 
calculated current scattered along the DNA axis and the square of the
wavefunction correlated well with the maxima found in the yield function for
breaking a single and two adjacent strands of DNA by the impact of 0-7 eV
electrons.  The correlation indicates that a substantial increase in DNA damage
occurs at the energies of preferential scattering of the incoming electron
along the DNA axis; i.e., at energies where increased coherent enhancement
occurs due to base stacking.  Although our results for A-DNA and B-DNA are not
markedly different, theory-experiment correlation appears to be better for the
A form, which is likely to be the one present in thin film experiments.\cite{PRA} It
appears from both comparisons, that diffraction effects within DNA should be
taken into account to describe the mechanism of action of electrons with
energies below 7 eV.  This may also hold true for higher energies, but it is
difficult to compare the present calculations with experimental data beyond 7
eV, because core-excited resonances implicated in the damaging process were not
included in our model and because the energy dependence of the
scattered current and of the square of the wavefunction change according to
sequence.

Our model constitutes an attempt to relate the gas phase and condensed phase
areas of the research in electron scattering from biomolecules. The model is
far from perfect since it has many assumptions. In particular it is difficult
to draw solid conclusions on radiation damage from this model, first of all
because we consider only elastic scattering events, and completely neglect the
motion of the nuclei, which would add a prohibitive new level of complexity.
Also, considering DNA as rigid is far too simplistic in solution but also to
some extent in thin films where at best the material can be considered
"amorphous" and structural water is present anyway. The motion of the electron
in the liquid should also be considered, and it would be best to use an electron
distribution taken from photoemission in a biological medium like water
instead of using a plane wave \cite{Allan:JES89}. One simple improvement to the
model could be to consider structural water, with positions taken possibly from
relevant DNA crystal structures, and we are planning to explore this
possibility to gain a better understanding of the role of
disorder and parasite scatterers.

\section*{Acknowledgments}
S.T. is supported by NSF and NSEC.
The work of C.H.G. has been supported partly by an allocation of NERSC supercomputing resources that were used to perform the R-matrix calculations, and in part by the U.S. Department of Energy, Office of Science.

\bibliography{mult_scat_DNA6a}

\end{spacing}
\end{document}